\documentclass{webofc}

\usepackage[varg]{txfonts}   
\usepackage{hyperref}
\usepackage{url}
\hypersetup{colorlinks=true,citecolor=blue,urlcolor=blue,linkcolor=blue}
\begin{document}
\title{Updated Astrophysical Equation-of-State Constraints on the Color-Superconducting Gap}
%
%

\author{\firstname{Aleksi} \lastname{Kurkela}\inst{1}\fnsep\thanks{\email{aleksi.kurkela@uis.no}} \and
        \firstname{Krishna} \lastname{Rajagopal}\inst{2}\fnsep\thanks{\email{krishna@mit.edu}} \and
        \firstname{Rachel} \lastname{Steinhorst}\inst{2}\fnsep\thanks{\email{rstein99@mit.edu}}
}

\institute{Faculty of Science and Technology, University of Stavanger, 4036 Stavanger, Norway
\and
           MIT Center for Theoretical Physics -- a Leinweber Institute, Massachusetts Institute of Technology, Cambridge, MA  02139, USA
          }

\abstract{We summarize and update using new NICER measurements the results of Ref.~\cite{Kurkela:2024xfh}, in which we used various 
astrophysical neutron-star observations
to set an upper bound on the CFL color-superconducting gap in a range of baryon chemical potentials $\mu_B \in [2.1,3.2]$, above those reached within neutron stars.
We also corroborate the ``reasonable" constraint from Ref.~\cite{Kurkela:2024xfh} on the maximum value of the color-superconducting gap by performing a new Bayesian analysis using a prior that extends a two-segment Gaussian process connecting the whole density range between CEFT and pQCD.   
}

\qquad\qquad\qquad\qquad\qquad\qquad\qquad\qquad\qquad\qquad\qquad\qquad\qquad\qquad\quad MIT-CTP/5909

\maketitle
Quarks in sufficiently dense and cold matter are expected to form Cooper pairs, a phenomenon known as color-superconductivity.  Though this pairing in principle enters as a subleading correction to the equation of  state (EoS), such a correction has often been ignored. Though the color superconducting gap $\Delta$ is small at weak coupling~\cite{Son:1998uk}, which would suggest that the correction to the EoS is small, the weak-coupling calculation only holds at asymptotically large $\mu$ and model estimates yield values of $\Delta$ ranging between 20 and 250~MeV, typically around 50-100~MeV~\cite{Alford:2007xm}. Absent a reliable calculation of $\Delta$, 
it is interesting 
instead to ask what empirical constraints exist from recent neutron star (NS) observations. 

At a baryon chemical potential of $\mu\sim2.6$~GeV, at which perturbative QCD (pQCD) is generally accepted to be reliable~\cite{Gorda:2023usm}, 
the form of pairing is color-flavor-locked (CFL), in which all three colors and flavors of quarks participate in the pairing. In our recent work~\cite{Kurkela:2024xfh}, we assume only that the CFL gap $\Delta$ is slowly varying with $\mu$, and attempt to empirically constrain its possible range of values. Taking inspiration from work using the pQCD EoS to constrain the NS EoS by imposing causality and thermodynamic consistency~\cite{Komoltsev:2021jzg}, we invert such an analysis and use constraints on the NS EoS known from astrophysical observations to constrain possible non-perturbative corrections to pQCD at typical matching values of $\mu$. 

Specifically, we take the baryon density and pressure $(n_H,p_H)$ at some $\mu_H$ to be the sum of the N$^2$LO pQCD prediction and the leading-order CFL correction, i.e., 
\begin{align}\label{eq:nshift}
    n_H&=n_{\rm pQCD}(\mu_H)+n_{\rm CFL}=n_{\rm pQCD} (\mu_H)+\frac{2}{3\pi^2}\Delta(\mu_H)^2\mu_H \\
    p_H&=p_{\rm pQCD}(\mu_H)+p_{\rm CFL}=p_{\rm pQCD}(\mu_H)+\frac{1}{3\pi^2}\Delta(\mu_H)^2\mu_H^2\,.\label{eq:pshift}
\end{align}
Then if the NS EoS goes through some point $(\mu_L,n_L,p_L)$, thermodynamic consistency requires 
\begin{equation}
    p_H = p_L+\int_{\mu_L}^{\mu_H} n(\mu) d\mu\,,
\end{equation}
 where $n(\mu)$ should imply a sound speed $ c_s^2 =\left(\frac{d\log n}{d\log \mu}\right)^{-1}$ no greater than $c_{s,\text{max}}^2=1$. (We might more generally require some maximum speed of sound $c_{s,\text{max}}^2$ less than 1.) This means that $p_H$ can be no larger than the maximum possible area which can be enclosed by $n(\mu)$ between $n_L$ and $n_H$ while maintaining the minimum possible slope $\frac{d\log n}{d\log \mu}\geq c_{s,\text{max}}^{-2}$. Using our expression for $p_H$, this equates to a constraint on the value of the CFL gap at $\mu_H$,\begin{equation}\label{eq:deltaconstraint}
      \Delta_{\rm max}(\mu_H)^2 \leq \frac{3\pi^2}{\mu_L^2} \Bigg[ \frac{n_{\rm pQCD}(\mu_H)}{2 \mu_H} \Bigl( \mu_H^2 - \mu^2_L \Bigr) 
    - \Bigl( p_{\rm pQCD}(\mu_H)- p_L \Bigr) \Bigg]
 \end{equation}
for $c_{s,\text{max}}^2=1$.

In practice, such a thermodynamic point ($\mu_L,n_L,p_L$) is not calculable directly or known precisely from NS observations, but can be inferred using a Bayesian analysis of EoS curves generated using a Gaussian Process (GP). We update the posterior distribution from Ref.~\cite{Gorda:2022jvk} which we used in Ref.~\cite{Kurkela:2024xfh} that infers the EoS incorporating the mass-measurements of PSR J0348+0432~\cite{Antoniadis:2013pzd} and PSR J1624--2230~\cite{Fonseca:2016tux}, the simultaneous mass and radius measurement of PSR J0740+6620 obtained using the NICER telescope~\cite{Miller:2021qha} as well as the tidal deformability measurement of GW170817 achieved by the LIGO/Virgo collaboration~\cite{LIGOScientific:2018hze}. Here, we replace the measurement of Ref.~\cite{Miller:2021qha} with a more recent measurement of the mass and radius of PSR J0740+6620~\cite{Dittmann:2024mbo} which includes additional NICER data, and additionally we use the recent determination of the mass (from MeerKAT data) and radius (from NICER data) of J0614--3329~\cite{Mauviard:2025dmd}. In addition, the electromagnetic counterpart of GW170817 is accounted for by assuming that the final merger product is a black hole. We then for each EoS generated in the GP have some likelihood $P(\text{EoS} | \text{data} )$ of that EoS given the observations. 

The likelihood of a given CFL gap $\Delta$ given the observations can be calculated as
\begin{align}
    P(\Delta | \text{data}) = \int_{\rm EoS} P(\Delta | \text{EoS} ) P(\text{EoS} | \text{data} )= \int_{\rm EoS} \frac{P(\text{EoS} | \Delta ) P(\Delta)}{P(\text{EoS})} P(\text{EoS} | \text{data})
\end{align}
where we take a uniform prior $P(\Delta)$ between 0 and 1 GeV. The likelihood function $P(\text{EoS} | \Delta )$ is taken to be 0 if $\Delta>\Delta_{\rm max}$ according to Eq.~\eqref{eq:deltaconstraint}, and 1 otherwise, where we additionally marginalize over the renormalization scale $X\in (1/2,2)$ to account for the uncertainty of the pQCD calculation.

\begin{figure}
\centering
\sidecaption
\includegraphics[width=7cm,clip]{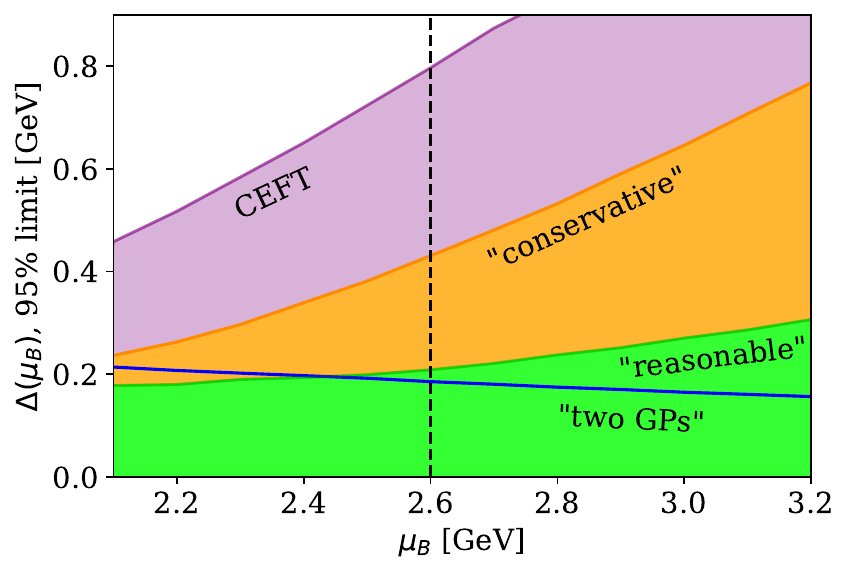}
\caption{95\% upper confidence bound on the CFL gap $\Delta$ at $\mu_B$. The CEFT, ``conservative", and ''reasonable" bounds are as in Ref.~\cite{Kurkela:2024xfh}, with the latter two updated to include  more recent NICER measurements. For these, we calculate the bound using the most extreme EoS above a certain density which respects a maximum speed of sound $c_{s,\text{max}}^2$. The ``two GPs" bound in blue uses both a high-density and low-density GP, and beyond requiring $c_s^2\leq 1$ in each generated EoS, no further limitation is made on $c_s^2$, unlike in the ``reasonable" scenario. We find that it yields a similar constraint to the ``reasonable" scenario.}
\label{fig:bound}       
\end{figure}

We consider three scenarios: First, a maximally conservative scenario in which we set $c_{s,\text{max}}^2=1$, and choose ($\mu_L,n_L,p_L$) for each EoS to as the central density/pressure of a 2.1~$M_\odot$ NS (or the maximum supported mass NS, only if this is smaller than 2.1~$M_\odot$), since this is approximately the mass of the most massive known NS. Second, a more reasonable scenario in which we set $c_{s,\text{max}}^2=1/2$ above the matching point, and choose ($\mu_L,n_L,p_L$) to always be the central density/pressure of the most massive NS supported by each EoS, even if this is larger than 2.1~$M_\odot$. Finally, for comparison, we choose ($\mu_L,n_L,p_L$) to correspond to the largest $\mu$ at which we trust CEFT ($\mu_L=0.97$~GeV,$n_L=1.1\,n_s$). The resulting 95\% bounds are shown in  orange, green, and purple in Fig.~\ref{fig:bound}, and are evidently very similar to those obtained in Ref.~\cite{Kurkela:2024xfh} before the newest NICER measurements were available. This is the first empirical constraint on the color-superconducting gap, and the similarity in order of magnitude of the ``reasonable" constraint to model predictions is striking.

The cutoff point, that is, the chemical potential at which we end each neutron star equation of state in the GP and switch to the most extreme equation of state which respects the maximum allowed $c_s^2$, has a non-trivial effect on the resulting constraint. Furthermore, this most extreme equation of state is somewhat unreasonable, as it discontinuously matches an EoS with the maximum allowed speed of sound to the pQCD speed of sound of approximately $c_s^2\approx 1/3$. In a sense, this is what the choice $c_{s,\text{max}}^2=1/2$ in our ``reasonable" scenario attempts to capture. The unreasonableness of the extreme intermediate EoS was pointed out in Ref.~\cite{Komoltsev:2023zor}, which presents an alternate approach in an effort to prefer a reasonable intermediate-density EoS, and ideally also reduce cutoff dependence: rather than choosing the most extreme possible high-density equation of state between each NS EoS and the pQCD EoS, one can introduce an additional GP conditioned by the pQCD EoS. Then at some fixed density, a matching is performed between GPs to require continuity of the energy density and pressure. We also note that another approach has since been demonstrated which uses only one GP and requires no specific choice of matching density~\cite{Finch:2025bao}. Here we will use the method introduced by Ref.~\cite{Komoltsev:2023zor} to 
corroborate the constraint on $\Delta$ we find in our ``reasonable" scenario without explicitly placing a constraint on $c_s^2$ beyond $c_s^2\leq 1$.

To do this, we begin with low- and high-density GPs conditioned on CEFT and pQCD respectively as obtained in Ref.~\cite{Komoltsev:2023zor}. That is, we will have an ensemble of EoSs extending to some fixed density $n_L$, and a second ensemble of EoSs extending from $n_L'=n_L-\frac{2}{3\pi^2}\Delta(\mu_H)^2\mu_H$ to ($\mu_H,n_{\rm pQCD},p_{\rm pQCD}$). Then instead of computing $P(\text{EoS} | \Delta )$ using Eq.~\eqref{eq:deltaconstraint} as above, we use the high-density ensemble to calculate a posterior distribution for ($\varepsilon (n_{\rm term}'),p(n_{\rm term}')$) by performing a kernel density estimation. We then compute $P(\text{EoS} | \Delta )$ by evaluating the kernel density estimation at ($\varepsilon_L-\varepsilon_{\rm CFL},p_L-p_{\rm CFL}$), effectively shifting the pQCD point ($\mu_H,n_{\rm pQCD},p_{\rm pQCD}$) up to ($\mu_H,n_H,p_H$) as in Eqs.~\eqref{eq:nshift} and~\eqref{eq:pshift}.

\begin{figure}
\centering
\sidecaption
\includegraphics[width=7cm,clip]{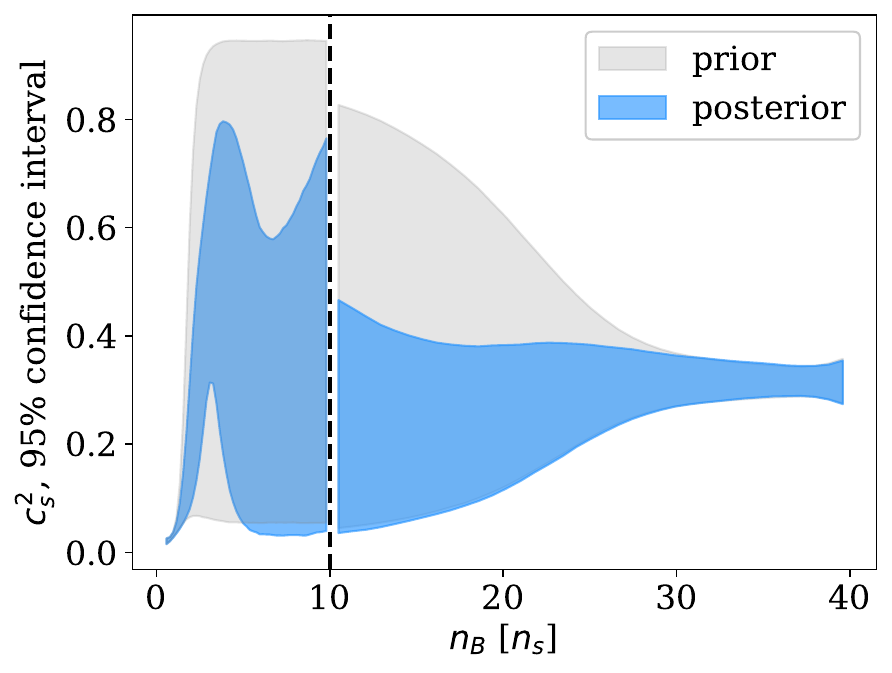}
\caption{95\% confidence interval for the speed of sound squared as a function of baryon density using separate low-density and high-density GPs. The low-density distribution is weighted by both likelihoods from the astrophysical measurements and a kernel density estimation of ($\varepsilon,p$) at $n=10n_s$ for the high-energy prior, while the high-density GP is weighted by a kernel density estimation of the astrophysically-weighted low-density GP at $n=10n_s$.}
\label{fig:cs2}       
\end{figure}

The result for $n_L=10n_s$ is shown in blue in Fig.~\ref{fig:bound}, a constraint very similar to the one we found in our ``reasonable" scenario. This makes some sense when we inspect the posterior distribution for the speed of sound (without gap) shown in Fig.~\ref{fig:cs2}; the high-density equation of state respects $c_s^2\leq 1/2$. It is also striking that the posteriors in this plot are nearly continuous, since we did not require continuity in the speed of sound when matching each low-density equation of state to the high-density ensemble.  We conclude from this comparison of different analyses that the assumptions made in our ``reasonable" bound on the gap are indeed reasonable.

Looking forward, the most promising avenue for a significant change in our bound on $\Delta$ is a more precise calculation of the high-density pressure. An increase to either the pQCD pressure or the CFL correction to the pressure at fixed $\mu$ and $\Delta$ would result in a stronger constraint. One such calculation has already been done since the publication of our paper---while we in Eqs.~\eqref{eq:nshift} and~\eqref{eq:pshift} consider only the leading order CFL contribution to the EoS (that is, $\mathcal{O}(g^0\Delta^2)$ where $g$ is the strong coupling), recently a calculation of the CFL contribution to the EoS to order $\mathcal{O}(g^2\Delta^2)$ has been performed \cite{Geissel:2024nmx}, finding a very large subleading term. This has been used to perform a similar analysis to ours, resulting in a stronger constraint on the CFL gap while at the same time posing the question of how the constraint could differ if contributions that are still higher order in $g$ are included~\cite{Geissel:2025vnp}. 
We also await the forthcoming full N$^3$LO pQCD calculation with great interest, as it is expected to reduce the uncertainty in $p_{pQCD}(\mu_H)$ considerably~\cite{Gorda:2023mkk}. If the updated $p_{pQCD}(\mu_H)$ were to land in the upper half of the current uncertainty band, this would tighten the constraint on the CFL gap.

\section*{Acknowledgements}

This research was supported in part by the U.S.~Department of Energy, Office of Science, Office of Nuclear Physics under grant contract number DE-SC0011090.  
KR acknowledges the hospitality of the CERN Theory Department and the Aspen Center for Physics, which is supported by National Science Foundation grant PHY-2210452.

\bibliography{main} 

\end{document}